\documentclass{aipproc}

\layoutstyle{8x11single}

\SetInternalRegister\hbadness{8000} 

\begin{document}

\title
      [Polarized Foregrounds Power Spectra vs CMB]
      {Polarized Foregrounds Power Spectra vs CMB}


\author{Carlo Baccigalupi}{
  address={SISSA/ISAS, Via Beirut 2-4, 34014 Trieste, Italy},
  email={bacci@sissa.it},
  thanks={This work was commissioned by the AIP}
}

\iftrue
\author{Gianfranco De Zotti}{
  address={Osservatorio Astronomico di Padova, Vicolo dell'Osservatorio 5, 35122 Padova, Italy},
  email={dezotti@pd.astro.it}
}

\iftrue
\author{Carlo Burigana}{
  address={ITeSRE-CNR, Via Gobetti, 101, I-40129 Bologna, Italy},
  email={burigana@tesre.bo.cnr.it}
}
\iftrue
\author{Francesca Perrotta}{
  address={Osservatorio Astronomico di Padova, Vicolo dell'Osservatorio 5, 35122 Padova, Italy},
  email={perrotta@sissa.it}
}

\copyrightyear  {2001}

\begin{abstract}
We briefly review our work about the polarized foreground
contamination of the Cosmic Microwave Background maps. We start by
summarizing the main properties of the polarized cosmological
signal, resulting in ``electric" (E) and ``magnetic" (B)
components of the polarization tensor field on the sky. Then we
describe our present understanding of sub-degree anisotropies from
Galactic synchrotron and from extra-Galactic point sources. We
discuss their contamination of the cosmological E and B modes.
\end{abstract}

\date{\today}

\maketitle

\section{Introduction}

Several ongoing or planned experiments are designed to reach the
sensitivities required to measure the expected linear polarization
of the Cosmic Microwave Background (CMB), see e.g. \cite{STAGGS}.
The forthcoming space missions MAP and {\sc Planck} aim at
obtaining full sky high resolution maps of Cosmic Microwave
Background (CMB) anisotropies, up to several arcminute resolution
(see e.g.\cite{MAP,PLANCK}; MAP webpage: {\tt
http://map.gsfc.nasa.gov/}; P{\sc lanck} webpage: {\tt
http://astro.estec.esa.nl/SA-general/Projects/}{\tt Planck/}).

They will also probe the polarization of the CMB radiation.
MAP has polarization sensitivity in all channels.
The current design of instruments for the {\sc Planck}
mission provides good sensitivity to polarization at all LFI (Low Frequency
Instrument) frequencies (30--100 GHz) as well as at three
HFI (High Frequency Instrument) frequencies (143, 217 and 545 GHz).

While there is a very strong scientific case for CMB polarization
measurements (cf., e.g., \cite{ZALDARRIAGA} and references therein), they
are very challenging both because of the weakness of the signal and
because of the contamination by foregrounds that may be more polarized
than the CMB.


In this paper, we begin by giving a description of the key
features and meaning of the polarized CMB component. The latter is
usually described in terms of the angular power spectra of two
components of the CMB polarization signal, namely electric (E) and
magnetic (B) modes (see \cite{HU} for an extensive treatment).
Then we summarize the main results of our recent work \cite{BACCI}
on the diffuse Galactic synchrotron polarized emission, focusing
on sub-degree anisotropies. These results have been obtained by
analyzing the existing high resolution data in the radio band: the
Parkes and Effelberg surveys at 2.4 and 2.7 GHz \cite{DUNCAN}
along the Galactic plane, and the medium latitude data at 1.4 GHz
\cite{UYANIKER}. Moreover we present some preliminary results on
the power spectrum of polarized emission from extragalactic radio
sources obtained exploiting data from the NVSS survey
(\cite{NVSS}; {\tt http://www.cv.nrao.edu/\~jcondon/nvss.html}).
In the last Section we give some concluding remarks.

\begin{figure}
\includegraphics[height=.5\textheight]{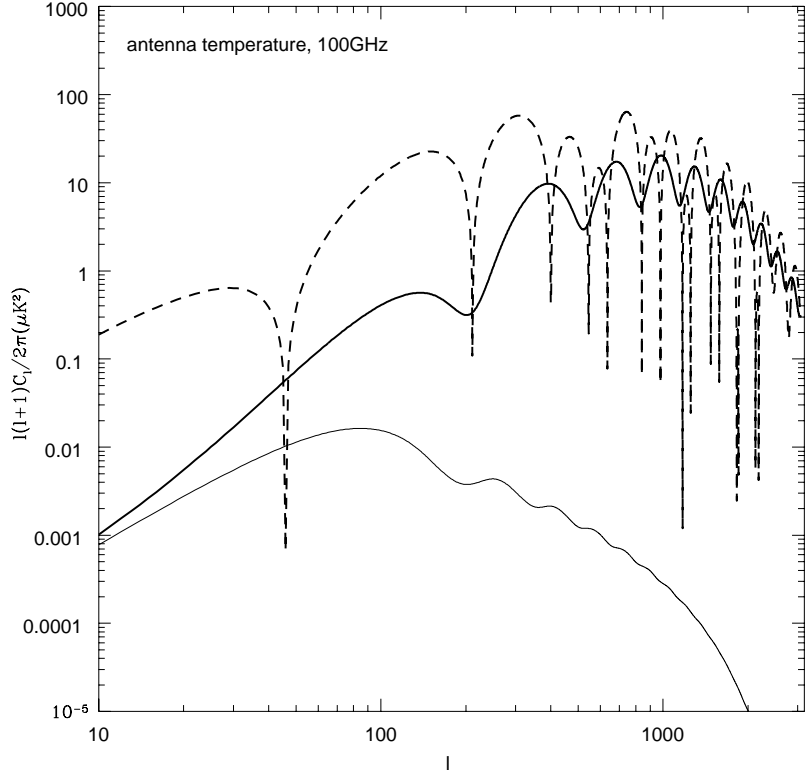}
\caption{Angular power spectrum of CMB polarization for the 
cosmological model described in the text. Dashed lines refer to
the TE correlation, light solid lines to B and heavy solid lines
to E modes.} 
\label{f1}
\end{figure}

\section{Cosmological polarisation modes: electric and magnetic type}

We give here only the basic features of the current description of
the cosmological polarized signal. A detailed treatment can be
found, e.g., in \cite{HU}.

Given two orthogonal axes $i,j$ in the plane perpendicular to the
photon propagation direction $\hat{n}$, the $2\times 2$ linear
polarization tensor $I_{ij}$ is represented by the Stokes
parameters $Q$ and $U$, with $Q=(I_{11}-I_{22})/4$, $U=I_{12}/2$.
It is convenient to define the complex quantities $Q\pm iU$, which
transform like a definite spin state under rotation by an angle
$\psi$ around $\hat{n}$:
\begin{equation}
\label{QUrotation}
(Q\pm iU)\rightarrow e^{\mp 2i\psi}(Q\pm iU)\ .
\end{equation}
These quantities can be expanded in the tensor spherical harmonics
$_{\pm 2}Y_{l}^{m}$ as
\begin{equation}
\label{QUexpanded} (Q\pm iU)(\hat{n})=\sum_{lm}a_{\pm
2,lm}\,{}_{\pm 2}Y_{lm}(\hat{n})\ .
\end{equation}
The expansion coefficients for E and B modes can then be defined
as:
\begin{equation}
\label{almEB}
a_{E,lm}=-(a_{+2,lm}+a_{-2,lm})/2\ ,\ a_{B,lm}=i(a_{+2,lm}-a_{-2,lm})/2\ .
\end{equation}
The electric and magnetic analogy comes from the properties of E
and B modes under parity transformation $\hat{n}\rightarrow
-\hat{n}$: while the $a_{E,lm}$ remain  unchanged, the $a_{B,lm}$
change sign \cite{HU}. The power spectra associated with E and B
modes, as well as their relation with the power spectrum of $Q$
and $U$ can be easily evaluated as:
\begin{equation}
\label{}
C_{l}^{E}={1\over 2l+1}\sum_{m}|a_{E,lm}|^{2}
\ ,\ C_{l}^{B}={1\over 2l+1}\sum_{m}|a_{B,lm}|^{2}
\ ,\ C_{l}^{E}+C_{l}^{B}={C_{l}^{Q}+C_{l}^{U}\over 2}\ .
\end{equation}
Due to the opposite parity properties, no correlation exists
between E and B. It is also useful to recall that E modes are
correlated with the total intensity fluctuations, giving rise to a
$C_{l}^{TE}$ power spectrum. The latter can be stronger than that
of  E and B CMB spectra since it receives contributions from total
intensity fluctuations that are expected to be 10 times or so
larger than the polarization ones. The description of the
polarization field in terms of E and B modes is more convenient
than the classical one in terms of local Q and U Stokes parameter
because while E receives contribution from all the types of
cosmological perturbations, B is non-zero only if vector or tensor
fields are present in the cosmological perturbations \cite{HU}. Of
particular interest are tensor perturbations associated to
gravitational waves because their amplitude is directly related to
the vacuum energy density during inflation.

To give a worked example, consider the constraints on cosmological
parameters set by the recent data from BOOMERanG \cite{BOOM}.
According to these data, the present cosmological energy density
is consistent with the critical one, being made by a $70\%$ of
vacuum energy ($\Omega_{\Lambda}=70\%$), a $25\%$ of dark matter
and a $5\%$ of baryons with an Hubble constant of 70 km/sec/Mpc
($\Omega_{b}h^{2}=0.022$). In Fig.~\ref{f1} the TE, E and B power
spectra are shown for this cosmological model, further assuming a
contribution to the temperature quadrupole of tensor perturbations
equal to 30\% of the contribution of scalar perturbations. The
main power resides in TE and E since these modes receive inputs
from acoustic oscillations occurring inside the horizon at
decoupling, corresponding to $l\ge 200$ in the figure. The B
component is subdominant since it is excited by gravitational
waves which decay rapidly inside the horizon.

This example gives an idea of the importance of measuring E and B
modes of the CMB polarization fluctuations. It is therefore
extremely important to study the power of the foregrounds as
contaminants to this signal. In the next two Sections we give our
present guess of the contamination coming from the low frequency
Galactic and extragalactic emissions, taking as reference model
the one presented in Fig.~\ref{f1}.

\begin{figure}
 \resizebox{30pc}{!}{\includegraphics{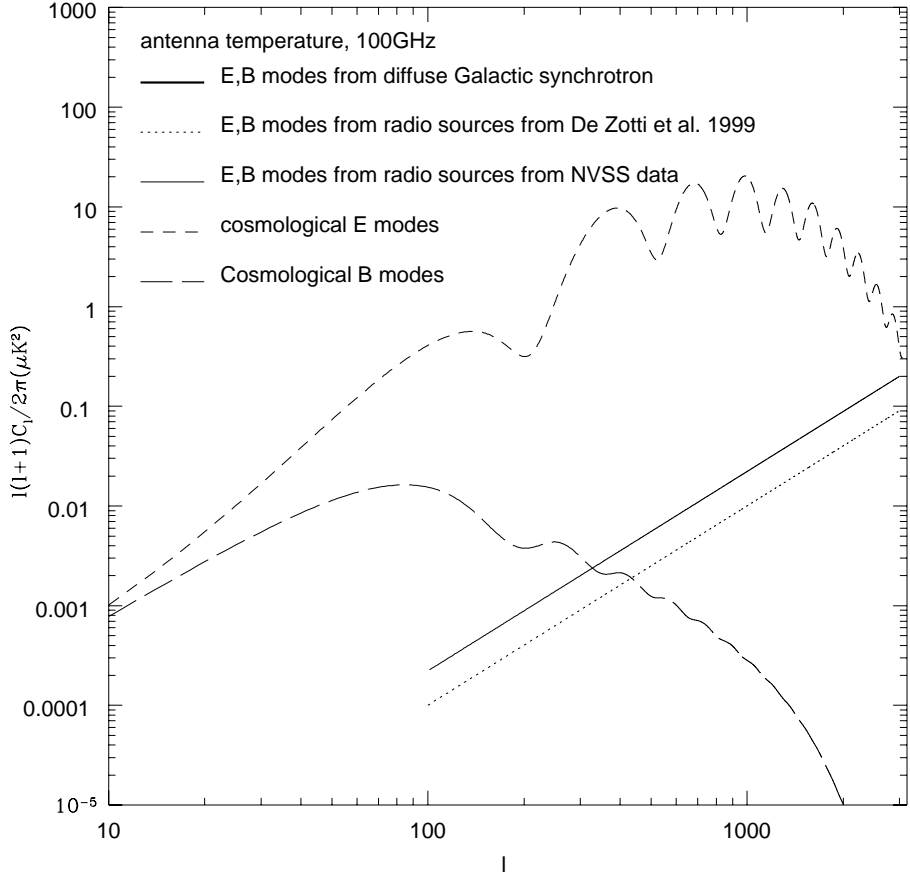}}
\caption{Comparison of cosmological E and B power spectra (short-
and long-dashed lines, respectively) with those from diffuse
synchrotron at low and medium Galactic latitudes (solid line at
$100\le l\le 1000$), and from extragalactic radio sources (solid
line at $100\le l\le 3000$) at 100 GHz.} \label{f2}
\end{figure}

\section{Polarized Galactic synchrotron emission}

We summarize here the main results of our recent paper
\cite{BACCI} in which we analyzed data from low and medium
Galactic latitudes at 1.4, 2.4, 2.7 GHz \cite{DUNCAN,UYANIKER},
having resolution of several arcminutes, and from high Galactic
latitudes on large angular scales \cite{BROUW}. We focus here on
the results concerning the power spectrum on sub-degree angular
scales, that we recast in terms of E and B modes.

By comparing total with polarized emissions we were able to
observe the following facts. The polarized emission does not show
any significant decrease with increasing Galactic latitude, up to
the highest latitudes considered ($|b|\simeq 20^{\circ}$), while
the total intensity decreases by a large factor. Correspondingly,
the polarization degree increases from typical values of a few
percent on the Galactic plane to about 30\% at latitudes
$10^{\circ}\le b\le 20^{\circ}$.

We found that the low polarization degree on the Galactic plane
can be largely explained by the contribution to the observed total
intensity from known intrinsically unpolarized HII regions,
catalogued by \cite{PALADINI}, which are concentrated on the
plane. We verified that, after removal of the contributions from
HII regions, the polarization degree drops to values consistent
with those found at medium latitudes. Of course, HII regions
themselves also contribute to Faraday depolarization of
synchrotron emission coming from outer Galactic regions.

Regions were identified where rotation measures towards pulsars
and extragalactic sources, the high polarization degree and, in
some cases, data on the distribution of polarization vectors and
on the Galactic magnetic field, consistently indicate low Faraday
depolarization.  The mean Galactic synchrotron power spectrum was
estimated as the average power spectrum of several such regions.
In terms of the E and B modes, and assuming a spectrum of the form
$S_\nu \propto \nu^{-0.9}$, i.e. antenna temperature $T_A\propto
\nu^{-2.9}$, we have, on degree and subdegree angular scales ($100\le l\le
1000$):
\begin{equation}
\label{clebsyn} C_{l}^{E}\simeq C_{l}^{B}=(1.2\pm 0.8)\cdot
10^{-9} \cdot\left(l\over 450\right)^{-1.8\pm 0.3}
\cdot\left(\nu\over 2.4{\rm\ GHz}\right)^{-5.8}\ \hbox{K}^2\ .
\end{equation}
The power is almost equally distributed among E and B modes, as is
expected since the alignment is preferentially determined by
magnetic fields, which do not have the characteristic parity
properties of scalar density perturbations (see \cite{SELJAK}).

In Fig.~\ref{f2} we plot this results (solid line at $100\le l\le
1000$), scaled to 100 GHz, against the different components of the
CMB spectrum shown in Fig.~\ref{f1}. This, albeit preliminary,
estimate, suggests that contamination from diffuse synchrotron is
not a serious hindrance for measuring the the CMB E-mode
polarization, but poses a serious challenge for measurements of
the B-mode power spectrum. 

\section{Extragalactic radio sources}

The confusion fluctuations due to a Poisson distribution of
extragalactic sources in the case of a polarimetric survey have
been discussed by \cite{SAZHIN} and \cite{DEZOTTI}. Briefly, in
the case of a population with uniform evolution properties and
constant (time-independent) polarization degree $\Pi$, the
polarization fluctuations $\sigma^2_P$ for cells of solid angle
$\omega$ are simply given by
\begin{equation}
\sigma^2_P = \sigma^2_I \langle \Pi^2 \rangle \ ,
\end{equation}
where $\sigma^2_I$ is the amplitude of intensity fluctuations for
the given cell size (see, e.g. \cite{TOFFOLATTI}) and
\begin{equation}
\langle \Pi^2 \rangle = \int_0^1 \Pi^2 p(\Pi) d\Pi\ ,
\end{equation}
$p(\Pi)$ being the distribution function of the polarization
degree. Clearly, an uncorrelated source distribution give equal
contributions to the E- and B-mode power spectra.

The estimates by \cite{DEZOTTI} exploited the models by
\cite{TOFFOLATTI} to estimate $\sigma^2_I$. To estimate the mean
polarization degree, they defined a complete sub-sample of BL-Lacs
for which polarization measurements at cm wavelengths are
available. The mean polarization degree at $\lambda = 2\,$cm was
found to be 5\%. The available data at shorter wavelengths suggest
that the polarization degree remains constant down to $\lambda
\simeq \,$ few mm. The E-mode (or B-mode) power spectrum of
polarization fluctuations due to radio sources, assuming $\Pi =
0.5\%$ for all populations contributing to the 100 GHz counts, is
shown in Fig.~\ref{f2}.

A new analysis, currently underway by \cite{MESA}, exploits the
NRAO VLA Sky Survey (NVSS) \cite{NVSS} which has provided I, Q,
and U data at 1.4 GHz for almost $2\times 10^6$ discrete sources
brighter than $s \simeq 2.5\,$mJy over about 10.3 sr of sky (about
82\% of the celestial sphere). Whenever possible, spectral indices
of sources have been determined combining the 1.4 GHz flux
densities with those given by the GB6 \cite{GB6} and PMN
\cite{PMN} catalogues at $\simeq 5\,$GHz. Extrapolations of
polarized fluxes to higher frequencies have been made assuming
that the polarization degree is frequency independent. A very
preliminary estimate of the derived polarization power spectrum is
shown in Fig.~\ref{f2}.

Advantages of this latter approach are that it automatically takes
into account the real space distribution of point sources,
including clustering effects, as well as the actual distribution
of their polarization properties. The large extrapolations in
frequency introduce, however, substantial uncertainties. On one
side, the polarization degree may be higher at higher frequencies
both because the Faraday depolarization becomes negligible and
because additional polarized components become optically thin. A
hint in this direction is provided by the fact that the mean
polarization degree of NVSS sources turns out to be $\simeq
1.4\%$, to be compared with the 5\% mean polarization at 15 GHz
found by \cite{DEZOTTI}. On the other side, it is known that many
sources with flat or inverted spectrum up to $\sim 5\,$GHz show
spectral breaks at higher frequencies. Thus, the assumption of a
constant spectral index up to 100 GHz leads to an overestimate of
polarization fluctuations. The two effects go in opposite
directions and therefore tend to counterbalance each other. To the
extent that the hypothesis of constant spectral indices holds, the
effective spectral index is found to be $\alpha_{\rm eff} \simeq
0.1$ ($S_{\nu}\propto \nu^{-\alpha}$, which becomes $T_{A}\propto 
\nu^{-2-\alpha}$ in antenna temperature). In order to estimate the
polarization fluctuation power spectrum we need to specify the
maximum flux of contributing sources (i.e. the minimum flux of
sources that can be individually detected and subtracted out).
Assuming that all sources with total flux larger than 5 times the
global rms fluctuations (including contributions of noise, CMB and
Galactic foregrounds), as estimated by \cite{TOFFOLATTI}, can be
removed, the E- and B-mode power spectrum of polarization
fluctuations due to extragalactic sources is described by:
\begin{equation}
\label{radiosourcesspectrum} C_{l}^{E}\simeq C_{l}^{B}=1.4^{+0.7}_{-0.4}
\cdot 10^{-7}\mu K^{2} \cdot\left({\nu\over 100\,{\rm
GHz}}\right)^{-4.2}\ ,
\end{equation}
This result is represented by the solid line extending up to
$l=3000$ in Fig.~\ref{f2}. We must caution that the assumption
about the flux limit for source subtraction may be somewhat
optimistic, so that the amplitude of fluctuations may be somewhat
underestimated. On the other hand it is reassuring that the two
totally independent estimates mentioned above give quite similar
results. It is also interesting to note that the power spectrum
derived from NVSS data is fully consistent with a Poisson
distribution of sources: clustering effects turn out to be
essentially negligible, as argued by \cite{TOFFOLATTI}.

\section{Conclusions}

There is growing interest and excitement about CMB polarization
studies. Measurements are extremely challenging because of the
extreme weakness of the signal to be detected. So, advances in
experimental techniques will be crucial, particularly to measure
the B-mode power spectrum, induced by gravitational waves. On the
other hand, it is not yet clear whether our ability to measure the
CMB polarization power spectrum  will be limited by detector
sensitivity or by foregrounds. In fact, polarized foregrounds are
currently very poorly understood.

On the other hand, new surveys are providing important pieces of
information, on which we can found preliminary but quantitative
estimates of the effect of foregrounds. We have focussed here on
polarized synchrotron emission from our own Galaxy and on
extragalactic radio sources. As for synchrotron emission, recent
high resolution and high sensitivity polarization maps at
frequencies in the range 1.4--2.7 GHz (\cite{DUNCAN,UYANIKER}),
although covering rather limited regions of the sky, have allowed
to estimate the power spectrum at sub-degree angular scales.

We have also presented and briefly discussed polarization
fluctuations due to extragalactic radio sources, based on two
approaches. On one side there are estimates based on counts as a
function of total flux, complemented with estimates of the mean
polarization degree. On the other side, the polarization
measurements provided by the NVSS were used together with
estimates of the spectral index of individual sources derived by
combining NVSS data with higher frequency catalogues (GB6 and
PMN). The two approaches yield results very close to each other.

Although the analysis is admittedly preliminary and does not
consider yet other potential polarized foregrounds at cm/mm
wavelengths (e.g. magnetic or spinning dust grains, see
\cite{DRAINE}), some indications are already emerging. Polarized
foregrounds do not seem to be a serious hindrance for measurements
of the CMB E-mode power spectrum on degree and sub-degree angular
scales, particularly in the frequency range 60--100 GHz (see
\cite{PRUNET} for a discussion of polarized foregrounds at higher
frequencies). However, foregrounds appear to be a potentially
serious limiting factor for experiments aimed at detecting B-mode
CMB polarization. More data and more detailed analyzes will
therefore be essential for designing future experiments.

\begin{theacknowledgments} 
We thank Dino Mesa for communicating results on extragalactic radio sources in 
advance of publication. Work supported in part by ASI and MIUR. 
\end{theacknowledgments}


\end{document}